\documentclass[11pt]{article}

\usepackage[T1]{fontenc}
\usepackage[utf8]{inputenc}
\usepackage{lmodern}
\usepackage{microtype}
\usepackage{amsmath}
\usepackage{graphicx}
\usepackage[hidelinks]{hyperref}

\title{Quantum-Annealing-Assisted Search-Space Reduction for\\
Large-Scale Shortest-Path Computation}

\author{%
  Kasymkhan Khubiyev\thanks{Corresponding author. Email: \href{mailto:hubiev.k@sirius.ru}{\texttt{hubiev.k@sirius.ru}}}
  \quad
  Yaroslav Kholodov\thanks{Email: \href{mailto:kholodov.ya@sirius.ru}{\texttt{kholodov.ya@sirius.ru}}}\\[0.3em]
  \normalsize Sirius University of Science and Technology\\[0.8em]
  Alexey S. Petrenko\thanks{Email: \href{mailto:a.petrenko1999@rambler.ru}{\texttt{a.petrenko1999@rambler.ru}}}\\[0.3em]
  \normalsize St. Petersburg Federal Research Center\\
  \normalsize of the Russian Academy of Sciences
}

\date{}

\begin{document}

\maketitle

\begingroup
\setlength{\leftmargini}{0pt}
\enlargethispage{2\baselineskip}
\begin{abstract}
Computing point-to-point shortest paths in large weighted graphs can require
classical algorithms to explore a substantial portion of the graph. We ask
whether quadratic unconstrained binary optimization (QUBO) can reduce the
work required to return a verified shortest path. For a known set of queries, we select paths
that share arcs. For new queries, we select a corridor and test whether
any path leaving it could improve the distance found by Dijkstra's algorithm.
We also partition the graph into blocks and reuse local paths between their
boundaries. We compare path encodings on identical inputs using simulated annealing
on classical processors.

The comparison of exact block overlays with direct Dijkstra search covered
45 synthetic graphs and the New York road network with 264,346 vertices.
The overlay reduced the number of arc inspections in the final search in
97 of 200 graph and block-size combinations. Once endpoint attachment,
assembly, and reconstruction were included, every combination took longer
per query than direct Dijkstra search. On a separate set of 168 synthetic
graphs, tightening the flow penalty increased the fraction of trials
returning a verified local optimum within 20\,ms from 35.8\% to 84.1\%
under the same annealing schedule and update policy. A cost model for a
prepared graph of 16,384 vertices predicts a $1.072\times$ query speedup
over a parallel classical overlay if five local problems can be solved
accurately in one batch at the assumed service cost. The practical benefit of reduction thus depends on whether savings in the
final search cover the cost of obtaining and verifying the local paths.
\end{abstract}
\endgroup

\section{Introduction}

A point-to-point routing query asks for a minimum-cost path between two
vertices of a weighted graph. In a road network the cost might be distance
or travel time. With nonnegative arc weights, Dijkstra's algorithm settles
vertices in order of increasing distance from the source until it reaches
the target \cite{dijkstra1959}. Even when the returned path is short, many
vertices off that path can be closer to the source than the target is.
The cost of discovering the path can therefore substantially exceed the
cost of traversing it.

Classical routing methods exploit different kinds of information to
restrict this search. The ALT algorithm uses A*, landmarks, and the triangle
inequality to bound the remaining cost to the target \cite{goldberg2005}, and contraction hierarchies (CH)
store shortcuts that can be reused across queries \cite{geisberger2008,dibbelt2016}.
When the pairs of interest are known in advance, a distance preserver can
instead retain a single subgraph containing a shortest path for each pair
\cite{bodwin2017,bodwin2021}. These methods suggest two criteria for a useful
reduction. It must preserve the distance required by the query, and the
search work it saves must cover the cost of preparing and using it.

Quadratic unconstrained binary optimization (QUBO) provides a way to
choose what to retain. Binary variables can select paths, regions, or arcs,
and quadratic terms can enforce consistency among those choices. Earlier
work used such models to partition graphs \cite{ushijima2017,hartmann2025}
and to select routes from a set of candidates \cite{neukart2017}. Here we
use QUBO decisions to prepare the graph on which an exact classical search
operates.

\emph{When does a smaller subgraph reduce the work of returning a verified
shortest path?} For a fixed set of demands, several routes may share arcs.
We compare an optimized union of routes with greedy selection from the
same candidate paths. For a new query, a route found inside a selected
corridor is optimal in the original graph only if no cheaper route leaves
the corridor. We compare corridor choices after charging for the boundary
checks and any additional search needed to establish that fact. We also
examine whether minimizing QUBO energy when choosing landmarks or ordering
vertices predicts the work later performed by the routing algorithm.

\emph{When do reusable paths inside graph blocks lower the cost of a
complete query?} A path between two boundary vertices can serve as one arc
in a smaller graph, called an overlay. Its usefulness depends on how many
such connections the blocks require and how much work remains to attach
the query endpoints. We compare direct Dijkstra search with overlays built
from exact local distances and from annealed local paths. Counting final
arc inspections alongside complete query time reveals where the reduction
saves work and where it moves work to another stage.

\emph{How does the encoding of a local path affect the chance of recovering
an optimum within a fixed budget?} We analyze the coefficient matrices of
alternative encodings, derive conditions under which their ground states
represent shortest paths, and compare sampled routes with exact reference
distances. The annealing experiments use classical simulated annealing on
CPUs. A separate cost model estimates the query speedup if independent
local problems can be solved concurrently on a quantum annealer, with the
static overlay already available to both the hybrid and classical methods.

The experiments on synthetic graphs and the New York road network show why
these questions must be considered together. Exact block overlays reduced
the final search on graphs with well-separated communities, yet the measured
cost of using the overlay outweighed that saving. A tighter flow penalty
made local optima more likely within a deadline. Whether that gain carries
through to a complete query depends on the cost of supplying accurate local
paths at the points where the global search needs them.

\section{Related Work}

\subsection{Classical shortest-path methods}

Dijkstra's algorithm \cite{dijkstra1959} computes
shortest paths from a source by repeatedly settling the unsettled vertex with the 
smallest tentative distance and relaxing its outgoing edges. 
A* \cite{hart1968} instead targets a particular
destination, ordering vertices by the sum of the distance already traveled and a 
heuristic estimate of the remaining distance. 
Thus, Dijkstra explores solely according to known path cost, whereas A* uses 
problem-specific guidance to focus the search; with an admissible heuristic 
it preserves optimality, and with a zero heuristic it reduces to Dijkstra's algorithm.

\subsubsection{Acceleration mechanisms}

Exact point-to-point speed-up techniques differ mainly in how much auxiliary 
information they compute and how that information is used during a query. 
At one end, query-only methods reduce exploration without preprocessing. 
Goal-directed and hierarchical methods use progressively richer information 
to guide or restrict graph search, while lookup-based methods move nearly all 
computation offline. Customizable methods introduce a separate trade-off: 
they preserve fast queries while allowing edge-cost metrics to change. 
The approaches below are therefore grouped by mechanism rather than strictly 
by publication date.

\paragraph{Query-only acceleration.}

Dantzig proposed running Dijkstra's algorithm simultaneously from the source and 
the target, alternating between the forward and backward searches
\cite{dantzig1963}.
His description in *Linear Programming and Extensions* was brief and did not fully 
specify the algorithm. Nicholson subsequently presented the first precise 
formulation \cite{nicholson1966},
dynamically expanding the direction whose next tentative distance is smaller and 
using a termination condition that guarantees optimality. 
Bidirectional search requires no preprocessing and often explores substantially 
less of the graph than one-sided Dijkstra, although its worst case may still examine 
the entire graph.

Despite its long-standing practical use, classical variants of bidirectional 
Dijkstra lacked a comparable general optimality guarantee. 
Haeupler et al. \cite{haeupler2025} close
this gap with a carefully balanced variant that alternates individual edge 
relaxations between the two directions. They prove that it is instance-optimal, 
up to a constant factor, in the number of examined edges when no heuristic or 
preprocessed information is available. The result applies to directed and undirected 
weighted multigraphs with positive edge weights and suitable access to edges from both 
search directions. It strengthens bidirectional Dijkstra as a preprocessing-free 
baseline, but does not establish superiority over ALT, CH, or other methods supplied 
with auxiliary information.

\paragraph{Guided search and pruning.}

Bidirectional Dijkstra reduces exploration using only information discovered during the query. Reach and ALT go further by preprocessing complementary kinds of auxiliary information: reach discards vertices using importance bounds, whereas ALT uses landmark-derived lower bounds to prioritize promising directions.

Reach assigns high importance to vertices that can occur far from both ends of
a shortest path \cite{gutman2004}. A query prunes a vertex when its reach bound
is too small relative to both the traveled distance and a lower bound to the
target. Reach needs only one extra value per vertex, remains exact with safe
upper bounds, and combines with A*. Tight bounds, however, are expensive to
compute; loose bounds prune little, and graphs without a strong hierarchy
benefit less.

Whereas reach decides which vertices may be safely discarded, ALT changes the
order in which the remaining vertices are explored. It selects landmarks,
precomputes distances to and from them, and uses triangle inequalities to
obtain admissible A* potentials \cite{goldberg2005}. Unlike Euclidean A*, it
needs no coordinates and works on directed graphs. It is simple, exact, and
compatible with bidirectional search, but consumes storage proportional to the
number of landmarks and depends strongly on landmark placement. Weak landmark
bounds may fail to beat bidirectional Dijkstra.

\paragraph{Metric-dependent hierarchies.}

Reach and ALT retain the original graph and decide much of the search online. Hierarchical methods instead encode the road network's structure during preprocessing and add shortcuts that allow queries to bypass less important regions.

Highway Hierarchies introduced a multilevel view in which local edges are used
near the endpoints and increasingly important subnetworks are used for longer
routes \cite{sanders2005}. Shortcuts preserve paths whose internal low-level
vertices are removed. This sharply reduces road-network search spaces and
established hierarchy as an alternative to goal-directed heuristics such as
ALT. Its preprocessing and query logic are comparatively complex, however, and
the hierarchy has no comparable advantage on arbitrary graphs lacking
road-like structure.

CH retains the hierarchical shortcut idea but replaces the more elaborate
level construction with a single vertex order. It assigns every vertex an
importance rank, contracts vertices from least to most important, and adds a
shortcut whenever no equally short witness path avoids the contracted vertex
\cite{geisberger2008}. Queries search upward in rank from both endpoints, then
unpack the selected shortcuts. CH is exact, fast, conceptually simpler than
Highway Hierarchies, and has modest auxiliary space. Its main weakness is
sensitivity to the contraction order and metric: poor orders generate many
shortcuts and large search spaces.

The later CH study developed the 2008 idea into a robust continental-scale
implementation and described up--down paths, conservative witness searches,
path unpacking, and practical ordering criteria in greater detail
\cite{geisberger2012}. It demonstrated that CH can support exact routing,
mobile queries, many-to-many search, and related extensions with small search
spaces. These are empirical road-network strengths rather than worst-case
guarantees, and distance-like or highly dynamic metrics can weaken the
hierarchy considerably.

\paragraph{Lookup-based routing.}

CH greatly reduces graph exploration but does not eliminate it. Transit node routing (TNR) and hub
labeling move farther along the preprocessing--query-time spectrum by storing
exact distances that can be combined at query time.

TNR assumes that every sufficiently long route passes through a small set of
transit nodes \cite{bast2007}. It stores all transit-node distances and a small
set of access nodes for each vertex, reducing nonlocal queries to table
lookups; local queries use a fallback search selected by a locality filter.
This can approach constant query time for long routes, but requires heavier
preprocessing and memory than CH, and its advantage largely disappears for
local routes.

Hub labels provide a more uniform lookup mechanism by moving nearly all work
into preprocessing \cite{abraham2011}. Each vertex stores hubs and exact
distances so that the source's forward label and target's reverse label share a
hub on a shortest path; a query only intersects the two sorted labels. Unlike
TNR, this mechanism does not distinguish local from nonlocal queries. In
contrast to ALT it stores exact witnesses rather than lower bounds, and in
contrast to reach and CH it replaces graph search rather than restricting it.
The benefit is extremely low, predictable query latency; the cost is
substantial preprocessing and memory, with difficult updates when weights
change.

\paragraph{Customizable routing.}

The fastest hierarchy- and lookup-based methods are usually optimized for a
particular edge-cost metric. CRP instead accommodates rapid changes in traffic,
turn costs, or user preferences by separating topology-dependent preprocessing
from metric-dependent customization.

Metric-independent preprocessing builds nested graph partitions, customization
computes exact boundary-to-boundary distances inside each cell, and a query
searches only relevant multilevel overlays \cite{delling2011}. Compared with
metric-dependent CH, CRP handles changing and personalized metrics much more
robustly and supports fast recustomization. Its static queries are usually
slower than those of a highly tuned CH, however, and its performance depends
on finding balanced partitions with small boundaries.

This dependence on partition quality motivates FlowCutter, which treats
separator construction as a trade-off between cut size and balance rather than
fixing one balance constraint \cite{hamann2016}. Better separators improve CRP
overlays and nested-dissection orders for customizable contraction hierarchy (CCH) preprocessing. The method
explores useful Pareto-optimal cuts efficiently and often improves downstream
routing structures, but remains heuristic in its terminal and piercing choices
and can spend substantial time computing an order. Unlike the preceding query
algorithms, it accelerates routing indirectly by improving the preprocessing
structure.

CCH combines the two main lines of development: it adopts CH-style upward
queries while using CRP's separation of topology and metric \cite{dibbelt2016}.
A metric-independent nested-dissection order creates a reusable chordal
supergraph; new weights are propagated through its triangles during fast
customization. Compared with CH, CCH sacrifices some static-query performance
for much faster metric changes; compared with CRP, it replaces multilevel
overlays with a contraction-based structure. Its quality depends on graph
separators, and triangle processing can require considerable time and memory.

\paragraph{Structural explanation and synthesis.}

The preceding techniques were initially motivated largely by their empirical success on road networks. Highway dimension provides a common structural explanation for that success, while the route-planning survey organizes their practical trade-offs.

Abraham et al. explain the observed success of reach, CH, TNR, and hub labels
through low highway dimension \cite{abraham2016}. Informally, a small local
hitting set intersects all sufficiently long shortest paths at each distance
scale. Multiscale hitting sets then yield landmarks or hubs, transit nodes,
hierarchy levels, shortcuts, and reach bounds with provable query and space
bounds. This unifies previously separate methods, but its strongest results
assume undirected graphs, positive lengths, unique shortest paths, and
expensive hitting-set preprocessing; the theoretical constructions are less
practical than engineered implementations.

These connections do not imply that one technique dominates the others. The
survey by Bast et al. places them on a common
preprocessing--space--query-time spectrum \cite{bast2016}: bidirectional search
and ALT favor low setup costs; reach and CH restrict graph search through
importance or hierarchy; CRP and CCH favor fast metric changes; and TNR and
hub labels buy near-table-lookup latency with more memory. The survey also
shows that shortcuts, labels, partitions, and landmarks are complementary
building blocks and that hybrids can outperform isolated techniques. Its
comparisons are necessarily approximate because implementations, hardware,
graph models, turn handling, and query distributions differ.

\subsection{Distance-preserving graph reduction}

Graph reduction can be performed before a shortest-path query by constructing
a \emph{distance preserver}: a subgraph that retains the exact distance for a
specified set of demand pairs. If the demands are all pairs in a terminal set,
the result is a subset preserver. Unlike query-time pruning or contraction
hierarchies, this approach removes edges from the underlying graph and
guarantees correctness only for the demands chosen during preprocessing.
Bodwin shows that even directed or weighted graphs admit preservers with
$O(n+n^{2/3}p)$ edges for $n$ vertices and $p$ demand pairs; in particular,
$p=O(n^{1/3})$ demands can always be preserved with only $O(n)$ edges
\cite{bodwin2017}.

The attainable sparsity depends strongly on graph type and on the demand set.
For undirected, unweighted graphs, consistent shortest-path trees and a
decomposition into induced matchings yield preservers with
$O(p+n^2/RS(n))$ edges, and hence $O(p)$ edges when
$p=\Omega(n^2/RS(n))$, where $RS(n)$ is the Ruzsa--Szemer\'edi extremal
function \cite{bodwin2021}. The proof techniques---consistent tie-breaking,
controlled branching, and delayed branching in shortest-path trees---identify
redundant edges without allowing any requested distance to increase.

These guarantees also expose limits of exact pruning. Some weighted
subset-preserver instances require $\Omega(n^{2/3}s)$ edges for $s$ terminals,
which is more than the number of terminal pairs when $s=o(n^{2/3})$; related
unweighted lower bounds are only slightly weaker \cite{bodwin2021}. Thus
aggressive reduction is possible when the query set is restricted or the
graph has favorable structure, but no uniformly tiny exact preserver exists
for every workload. Moreover, these results are principally existential and
do not provide a practical construction or empirical routing evaluation,
leaving room for heuristic or optimization-based preprocessing that searches
for sparse subgraphs while explicitly enforcing distance preservation.

\subsection{QUBO graph optimization and quantum annealing}

Quantum annealing addresses NP-hard graph problems by expressing their
objectives and constraints as quadratic unconstrained binary optimization
(QUBO), or equivalently Ising, energy functions. Ushijima-Mwesigwa et al.
formulate community detection, balanced bisection, and balanced concurrent
$k$-way partitioning in this form \cite{ushijima2017}. Modularity maximization
maps directly to an Ising quadratic form, while balance and one-hot assignment
constraints are introduced through squared penalties. Their main contribution
is methodological: the work shows how several graph-partitioning variants can
be represented on an annealer and introduces hardware-aware transformations,
including thresholding the dense modularity matrix and embedding a dense graph
through its sparse complement.

The reported partitions are often competitive in cut quality with METIS and
KaHIP, and the concurrent encoding can compute all $k$ parts at once rather
than through recursive bisection. These benefits come with substantial costs.
The one-hot construction requires $nk$ logical variables, penalty weights must
be large enough to enforce feasibility without overwhelming the useful
objective coefficients, and minor embedding can consume many physical qubits.
Moderate modularity thresholding preserves solution quality, but aggressive
thresholding degrades it. Moreover, the larger experiments rely on the hybrid
\texttt{qbsolv} decomposition and report solution quality without matched
end-to-end runtime or success-probability comparisons, so they demonstrate
useful formulations rather than quantum speedup.

Hartmann et al. specialize the same QUBO/annealing paradigm to two-way
power-grid partitioning for parallel simulation \cite{hartmann2025}. Instead
of minimizing a generic balanced cut, their objective combines estimated
component-model workload, sequential cut overhead, and imbalance in
network-solution work, with coefficients derived from floating-point operation
estimates. This makes the formulation more application-aware than that of
Ushijima-Mwesigwa et al.; it also treats balance as a soft performance
objective rather than solely as a feasibility constraint. In contrast, it is
restricted to two parts, does not require either part to be connected, and its
squared global balance terms make the logical QUBO fully connected even though
the underlying grid is sparse.

The two studies consequently expose different bottlenecks. Ushijima-Mwesigwa et al. emphasize encoding flexibility, sparsification, complement embeddings, and hybrid decomposition across general graph instances, whereas Hartmann et al. examine embedding choice, chain strength, anneal time, coefficient scaling, and time to solution on small grid cases against exact Gurobi results. Hartmann et al. find the optimum reliably only for the smallest instance; the 30-bus optimum is rare, the 57-bus optimum is not observed, and substantially larger QUBOs cannot be embedded in reasonable time. Objective rescaling can greatly improve sampling on one instance, but may change the optimization trade-off elsewhere. Thus both works identify QUBO design as valuable for encoding graph objectives, while also showing that dense logical interactions, physical embedding, parameter sensitivity, and incomplete end-to-end benchmarking remain central barriers to claiming practical quantum advantage.

In cybersecurity, the role of graph reduction depends on which network
properties the application must preserve. Khubiev et al.
review QUBO formulations for network defense, anomaly detection,
cryptanalysis, and cyber-risk assessment \cite{khubiev2026}. Their
discussion of network segmentation and sensor placement illustrates how
the choice of objective changes the meaning of a good solution. A partition
may separate security zones, while a placement is judged by the network
activity its sensors can observe. The review also distinguishes results
obtained with physical annealers, hybrid solvers, and classical heuristics.
For the FLS multigraphs motivating the present work, the relevant question
is whether a reduction preserves the minimum cost of an admissible
information-flow path between specified modules.

A direct precursor to the present study is the evaluation of path models
on security graphs by Khubiev et al.
\cite{khubiev2026action}. They compared an edge-flow encoding with a
variable-length sequence of hops, using classical heuristics and exact
reference solutions. Representing several actions for the same transition
increased model size, particularly for the hop encoding. A defensive
extension used the path costs to evaluate patch selections under a budget.
The present study carries the comparison of encodings into local path
problems. We examine whether reusing their solutions in a smaller graph
can reduce the cost of answering a complete query while preserving a
verifiable shortest path.

\subsection{Quantum optimization for network routing}

Quantum optimization studies related to source-to-target routing address
several distinct problems rather than a single quantum analogue of Dijkstra's
algorithm. Krauss et al. formulate maximum flow as QUBOs using parallel
unit-capacity edges, binary-encoded edge flows, and the max-flow/min-cut dual
\cite{krauss2020}. The direct-flow formulations return an explicit flow, but
conservation constraints create dense couplings and costly minor embeddings;
binary encoding reduces the logical-qubit count but not this connectivity
problem. The min-cut formulation is much smaller---one variable per vertex,
sparse interactions inherited largely from the graph, and support for real
capacities---but returns only the cut and maximum-flow value. It was
correspondingly the most successful formulation on the tested D-Wave hardware.
The experiments remain small, however, and their complexity discussion omits
embedding, programming, sampling, and validation costs, so it does not
demonstrate an end-to-end quantum speedup.

Neukart et al. instead optimize traffic by choosing one of at most three
classically precomputed routes for each of 418 cars \cite{neukart2017}. Their
1,254-variable QUBO minimizes squared road-segment occupancy, thereby
coordinating route choices so that congestion is not merely shifted from one
street to another. Hybrid \texttt{qbsolv} runs improved the reported congestion
measure over the original and random assignments, but the full problem was not
embedded directly on the QPU and the quantum contribution was not isolated
from tabu-search decomposition. Moreover, the objective is a static congestion
surrogate rather than shortest travel time: it omits time-dependent arrivals,
capacities, queues, and individual detour costs, and the study provides neither
a strong classical optimization baseline nor evidence of optimality or
scalability.

Harwood et al. consider the richer vehicle-routing problem with time windows
and compare route-, time-expanded arc-, and sequence-based QUBOs, plus a
continuous-time sequence model solved by hybrid ADMM \cite{harwood2021}. Their
central result is that formulation size alone is insufficient: precomputed
feasible routes give the smallest but densest QUBO and move substantial effort
into classical route generation; arc models represent timing and waiting more
explicitly but require the most variables; sequence models are sparser but
approximate timing unless continuous variables and a classical nonlinear
subproblem are added. Feasible bit strings become extremely rare as instances
grow, while exact penalty weights, circuit optimization, and solver performance
are highly formulation-dependent. Results on a noiseless 16-qubit simulator
show modeling feasibility and some tolerance to inexact QUBO subsolves, not
performance on realistic instances or quantum advantage.

The works therefore differ principally in what is optimized and where classical computation enters: Krauss et al. solve a single-commodity capacity problem, Neukart et al. select among fixed paths for many vehicles to balance congestion, and Harwood et al. construct complete constrained vehicle routes. None directly computes an unconstrained point-to-point shortest path on an arbitrary weighted graph. Across all three, the recurring barriers are logical-variable growth, dense penalty-induced interactions, minor-embedding or circuit-connectivity overhead, delicate coefficient scaling, and incomplete end-to-end comparison with strong classical methods. They nevertheless identify a useful design lesson for shortest-path QUBOs: preserving graph sparsity and reducing the candidate graph before encoding may matter as much as reducing the nominal number of binary variables.

\subsection{Large-QUBO execution strategies}

Bass et al. address QUBOs that are too large to embed directly by comparing
four hybrid decomposition methods on a D-Wave 2000Q \cite{bass2021}. Their
principal-component decomposition cuts a graph into independently annealed
pieces; Freeze-and-Anneal fixes variables predicted by a genetic algorithm;
Iterative Centrality Halo repeatedly solves a high-degree vertex and its
neighborhood; and \texttt{qbsolv} combines iterative sub-QUBO search with tabu
refinement. \texttt{qbsolv} gives the best solution quality overall, with ICH
generally second, suggesting that propagating solved subproblem decisions into
the remaining QUBO is more effective than solving pieces independently. The
study also identifies repeated minor-embedding searches as a major runtime
bottleneck: reusing a fixed embedding, as \texttt{qbsolv} does, can matter as
much as the decomposition rule itself.

Pelofske et al. consider the complementary case in which a QUBO is small
enough that a single embedding leaves much of the annealer idle
\cite{pelofske2022}. Their parallel quantum annealing method places multiple
disjoint clique embeddings on one chip and solves either several independent
maximum-clique QUBOs or several replicas of one hard QUBO in the same annealing
cycle. Although parallel execution slightly reduces the ground-state
probability of individual problems, amortizing a device call across many
embeddings substantially improves throughput; replicating one problem can
likewise improve its aggregate success probability. The benefit is largest
for small instances, since embedding capacity falls as QUBO size grows, and
it is sensitive to chain length, placement, cross-talk, unembedding cost, and
coefficient normalization when heterogeneous problems are combined.

The main distinction is therefore that Bass et al. decompose one oversized, interacting QUBO into sequentially coordinated subproblems, whereas Pelofske et al. compose mutually independent QUBOs into a block-diagonal problem for simultaneous execution. Decomposition expands the effective logical problem size but risks losing cross-subproblem information and spending excessive time on embedding; parallel annealing improves utilization and throughput but does not make any component QUBO larger or intrinsically easier. Both studies are empirical and hardware-dependent: Bass et al. do not compare against strong classical optimizers, while Pelofske et al. find that an exact classical clique solver remains superior in important sparse and larger-instance regimes. Consequently, neither establishes general quantum advantage, and both highlight embedding quality, classical overhead, parameter tuning, and matched end-to-end benchmarking as unresolved issues.

\section{Methodology}
\label{sec:methodology}

Let $G=(V,E,w)$ be a finite directed graph with positive arc weights. Each
arc has its own identity, so parallel arcs may have different costs. For
distinct vertices $s$ and $t$, write $d_G(s,t)$ for the minimum sum of arc
weights along an $s$--$t$ path, with $d_G(s,t)=+\infty$ if $t$ is
unreachable. Deleting arcs can only increase this distance. A reduction for
a reachable pair must therefore retain a path whose cost meets the required
bound. We consider two settings that differ in when the endpoints become
known. A fixed set of demands permits several routes to share retained arcs.
For a new query, a partition can be reused, but the selected regions must be
checked for that pair. We also examine a block overlay in which an arc can
represent an entire local path.

\subsection{Selecting a subgraph for a fixed set of demands}

Let $\mathcal D$ be a set of reachable ordered pairs. For each
$d=(s,t)\in\mathcal D$, we construct a nonempty set $\mathcal P_d$ of
candidate paths and check their costs against the full-graph distance.
Every admitted path $p$ satisfies $w(p)\leq(1+\varepsilon)d_G(s,t)$, where
$w(p)=\sum_{e\in p}w(e)$ and $\varepsilon\geq 0$ is the prescribed relative
tolerance. At $\varepsilon=0$, paths of equal minimum cost can still differ
in how many arcs they share with routes for other demands.

Let $x_e$ indicate retention of arc $e$ and let $z_{dp}$ indicate selection
of candidate $p$ for demand $d$. The nonnegative integer $c_e$ is the cost
of retaining $e$. Setting $c_e=1$ makes the objective count retained arcs,
irrespective of their traversal weights. We minimize
\begin{equation}
\label{eq:demand-union}
\begin{aligned}
 F(x,z)={}&\sum_{e\in E}c_e x_e
 +\lambda\sum_{d\in\mathcal D}
       \left(1-\sum_{p\in\mathcal P_d}z_{dp}\right)^2\\
 &+\lambda\sum_{d\in\mathcal D}\sum_{p\in\mathcal P_d}
                  \sum_{e\in p}z_{dp}(1-x_e),
 \qquad \lambda=1+\sum_{e\in E}c_e .
\end{aligned}
\end{equation}
The first penalty enforces one path per demand. The second requires every
arc of that path to be present, while the objective charges a shared arc
only once. Any violation costs more than retaining all arcs, so every global
minimizer is feasible. More importantly for heuristic optimization, \emph{any}
feasible assignment gives a distance guarantee. Its subgraph $G'$ contains
a validated path for each demand, hence
\[
 d_G(s,t)\leq d_{G'}(s,t)\leq(1+\varepsilon)d_G(s,t)
 \qquad ((s,t)\in\mathcal D).
\]
Optimization determines the cost of the union within the available candidate
pool. Constructing and validating that pool is part of the cost of serving
the declared demands. A change in demands requires a new assessment.

\subsection{Selecting and certifying a corridor for a new query}

We partition $V$ into disjoint regions and form a directed quotient graph.
A quotient arc joins two regions when an original arc crosses from the first
to the second. Each region receives a positive work estimate equal to one
plus its internal arc count, its outgoing boundary arc count, and the mean
number of scans attributed to it on training queries. Multiplication by the
number of training queries gives integer costs with the same ordering of
candidate corridors by total cost.

The query QUBO minimizes the sum of the costs of the retained regions.
Binary variables select regions and quotient arcs. Squared penalties require
both endpoint regions to be present and impose one unit of directed flow
from the source region to the target region. For each selected quotient arc,
a further penalty requires both incident regions to be retained. Each penalty
has coefficient one plus the sum of all integer region costs. If the
endpoints occupy the same region, that region provides the initial corridor.
For distinct endpoint regions, even a feasible quotient flow may enter and
leave an intermediate region at vertices that have no directed connection
inside it. We therefore test the corridor on the original graph.

Let $W$ contain all vertices of the selected regions. We compute
$U=d_{G[W]}(s,t)$ by Dijkstra's algorithm on the induced subgraph
\cite{dijkstra1959}. To decide whether a route outside $W$ could improve
this value, we use directed landmark bounds \cite{goldberg2005}.
The potentials $\ell_s$ and $\ell_t$ have $\ell_s(s)=\ell_t(t)=0$.
We verify on every original arc $e=(u,v)$ that
\[
 \ell_s(v)\leq\ell_s(u)+w(e),\qquad
 \ell_t(u)\leq w(e)+\ell_t(v).
\]
Summing along a path gives $\ell_s(u)\leq d_G(s,u)$ and
$\ell_t(v)\leq d_G(v,t)$. Let $\delta^+(W)$ contain every arc whose tail
is in $W$ and whose head lies outside it. We accept a finite $U$ if
\begin{equation}
\label{eq:corridor-certificate}
 U\leq \min_{e=(u,v)\in\delta^+(W)}
            \bigl\{\ell_s(u)+w(e)+\ell_t(v)\bigr\},
\end{equation}
with the minimum over an empty boundary taken as $+\infty$. Every path
leaving $W$ has a first outgoing arc, and the corresponding expression
bounds its cost from below, even if the path later re-enters $W$. Every path
remaining in $W$ costs at least $U$. Together these facts certify
$U=d_G(s,t)$, including when an equally short path leaves the corridor.

The verifier uses exact integer arithmetic for integer-weight graphs. If the
initial sample is infeasible, verification starts from the endpoint regions.
If the restricted search finds no path or the boundary test is inconclusive,
we add adjacent regions in order of increasing boundary bound and repeat the
search. After at most four expansion rounds, an unresolved query passes to
full-graph Dijkstra. We record where each query was resolved so that an exact
answer obtained after expansion or fallback can be distinguished from an
initially certified corridor.

\subsection{Routing through local shortest-path overlays}

A partition also permits repeated queries to reuse paths within blocks.
In each block, a boundary vertex is incident to an interblock arc in either
direction. We retain the original interblock arcs and construct a directed
overlay arc for each reachable ordered pair of distinct boundary vertices in
the same block. Each overlay arc stores a local path and its cost. At query
time we attach the source to reachable boundaries in its block and connect
boundaries in the target block to the target. When both endpoints lie in one
block, we also consider a direct local connection. Searching this overlay
still allows a route to leave that block and return.

\begin{figure}[htbp]
\centering
\includegraphics[width=\linewidth]{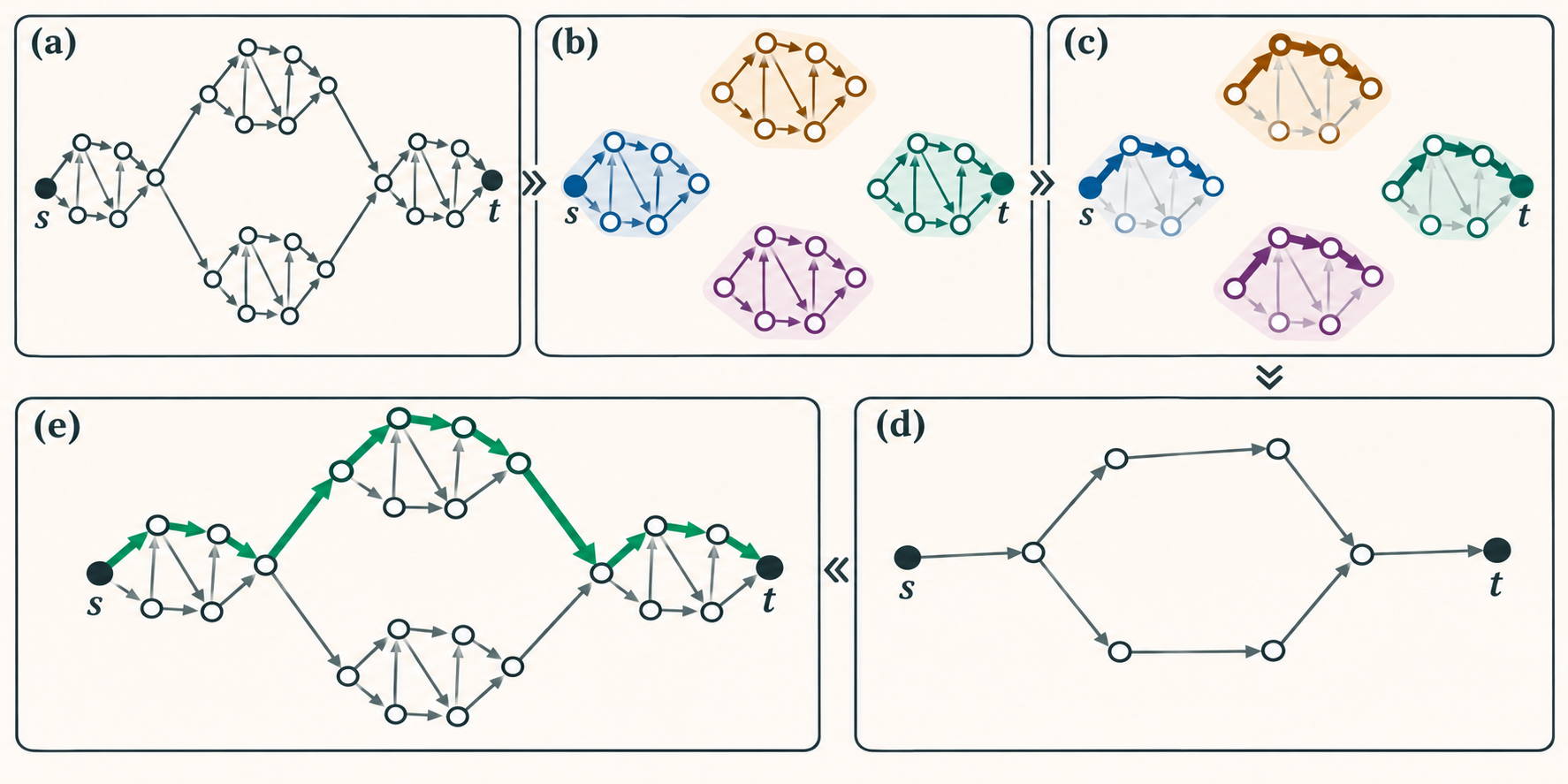}
\caption{Hybrid routing from $s$ to $t$. The input graph (a) splits into four blocks (b). The independent local problems (c) admit parallel quantum annealing. Bold arrows trace the selected local paths. Assembly (d) replaces each path with an arc of equal cost and restores the interblock arcs. Dijkstra selects a route on this overlay. Its expansion on the input graph (e) appears in green. The experiments used CPU simulated annealing for the local problems.}
\label{fig:overlay-algorithm}
\end{figure}

Figure~\ref{fig:overlay-algorithm} illustrates how the local paths form an answer to one query.

With exact local distances, the overlay preserves $d_G(s,t)$. Splitting an
original shortest path at its interblock crossings gives an overlay walk of
at most the same cost. Conversely, every overlay walk expands into original
arcs of exactly its stated cost. With heuristic local solutions, expansion
still verifies the cost of a returned route, but a local transition can be
too expensive or absent. We compare the heuristic overlay with both an exact
overlay built by local Dijkstra searches and direct full-graph Dijkstra. A
further control runs full-graph Dijkstra on every heuristic-overlay query
and includes that check in its query time.

The size of an overlay depends on its boundaries. For $m_{\rm cross}$
interblock arcs and $b_i$ boundary vertices in block $i$, it contains at most
$m_{\rm cross}+\sum_i b_i(b_i-1)$ arcs before endpoint attachment. We
therefore measure boundary storage, local preparation, endpoint attachment,
and final search separately. If one block contains the entire graph, its
endpoint attachment can already solve the original query.

\subsection{Encoding and verifying local paths}

For a reachable pair $a,b$ in a loop-free block with positive integer
weights, let $B$ be the out-minus-in vertex--arc incidence matrix. Let $h$
have entries $+1$ at $a$, $-1$ at $b$, and zero elsewhere. With one bit
$x_e$ per internal arc, the local flow model minimizes
\begin{equation}
\label{eq:local-flow}
 E_A(x)=\sum_e w(e)x_e+A\lVert Bx-h\rVert_2^2 .
\end{equation}
A feasible integral unit flow decomposes into an $a$--$b$ path and
possibly directed cycles. Positive weights exclude cycles from a minimum.
For an infeasible assignment, the integer residual $Bx-h$ sums to zero,
so its squared norm is at least two. The empty assignment attains that
bound at zero traversal cost. Consequently, if $C^*$ is the local
shortest-path cost, $A>C^*/2$ excludes all infeasible ground states,
whereas $A=C^*/2$ admits an infeasible tie. We compare the sufficient
choice $A=1+\sum_e w(e)$ with $A=\lfloor U_{\rm BFS}/2\rfloor+1$, where
$U_{\rm BFS}$ is the weighted cost of a path found by breadth-first search.
The latter supplies a valid penalty before $C^*$ is known.

We also examine arc encodings with degree penalties, auxiliary bits,
one redundant flow equation removed, or costs modified by vertex
potentials. Position-based encodings represent the sequence of vertices
using one-hot or domain-wall variables. They identify a representative arc
when parallel arcs of equal weight realize the same step. To test these
formulations, we enumerate all loop-free directed graphs on two or three
labeled vertices with each possible arc absent or weighted one or two.
For eligible models of at most 20 bits, we compare matrix energies with an
independent evaluation of the polynomial and check the decoded ground states
against reference shortest paths. Additional cases exercise parallel arcs
and tied routes. In comparisons of sampled encodings, success requires the
entire assignment to describe a simple optimal path, including consistent
auxiliary bits where required. An optimal path accompanied by an extraneous
cycle fails this test.

\subsection{Solver comparisons and evaluation}

The binary models use integer coefficients and store each pairwise term
once. The substitution $x_j=(1+\sigma_j)/2$, $\sigma_j\in\{-1,1\}$, gives
the corresponding Ising energy for quantum annealing. All measured solver
runs in this study used classical processors. We compared simulated
annealing and tabu search on common reduction models, rescored their
assignments with the original coefficients, and recorded feasibility before
repair. For the demand union, a greedy control selects paths by their
incremental retention cost. For corridors, classical quotient initializers
use the same certification and recovery procedure. Full-graph controls
include Dijkstra, bidirectional search, and landmark-guided A*. These
comparisons distinguish the contribution of the reduction objective from
the cost of finding and verifying a particular assignment.

We evaluated overlays on grids, directed rings with chords, and graphs with
community structure. The paired classical comparison used five graphs per
family at each of 16, 64, and 256 vertices, with 40 common queries per graph,
and 256 queries on the directed New York road-distance graph. For the local
encoding experiments, we compared annealing at fixed numbers of sweeps,
fixed ceilings on proposed updates, and fixed elapsed times. Under a time
limit, success required construction, sampling, and strict verification to
finish before the deadline. A trial in which construction consumed the
budget counted as a failure. We varied the temperature schedule and the
variable update order on fresh synthetic graphs and induced road regions.
For each road region the reference distance was computed within the induced
graph.

Complementary experiments examined how QUBO choices affect classical routing
structures. We compared balanced separator objectives by counting directed
triangle relaxations during hierarchy customization, landmark selections by
counting A* arc inspections, multiscale covers by counting stored entries,
and contraction batches by measuring query inspections. For covers we also
checked the local density and whether every required shortest path was hit.
A successful fallback query alone cannot establish cover validity. A
separate separator comparison used 200 previously unseen ten-vertex
ring-with-chords graphs with at most 23 directed arcs and a feasible balanced
root separator. Its target was a 5\% reduction in customization attempts
relative to minimizing separator size, conditional on that graph population.

For each routing method we record retained vertices and arcs, settled
vertices, arc inspections, and correctness against full-graph distances.
For approximate routes we report reachability and, for a returned route to
a reachable target, the ratio of its cost to the reference distance.
Timings include candidate or model construction, solving, materialization,
verification, expansion, and fallback as applicable. If $S$ is the measured
setup time and $\bar t_q$ the mean complete online time, the cost of serving
$J$ queries under fixed weights is
\[
 T(J)=S+J\bar t_q .
\]
We compare methods on the same queries and report setup separately so that
the effect of reuse is visible. Demand-specific candidate generation enters
the setup cost of a path union. Partition fitting and reusable boundary
paths enter setup for the corresponding methods, while query-specific
corridor construction and endpoint attachment enter online time.

For a static graph with repeated queries we also consider latency after
preprocessing. Given planar coordinates, a grid origin, and a cell width,
two coordinate divisions and floors identify a vertex's cell in constant
time. We assume that the boundary paths and the static part of the overlay
are already available to both methods. This accounting removes partition
construction and static assembly from each query. It retains endpoint
attachments, the overlay search, path reconstruction, and the checks needed
to return a route of the required quality.

Let $C$ denote the common online cost after separating out local endpoint
solves, and let $D$ denote the online cost of the chosen classical baseline.
Suppose that $a$ independent local problems fit on a quantum annealer in
batches of up to $p$ simultaneous embeddings \cite{pelofske2022}. If each
batch costs $\lambda$ at the required route quality, the modeled speedup
per query relative to that baseline is
\begin{equation}
 \rho(p)=\frac{D}{C+\lceil a/p\rceil\lambda}.
 \label{eq:parallel-speedup}
\end{equation}
Here $\lambda$ includes per-batch communication, programming, repeated
reads, decoding, and any additional verification or recovery needed for
that quality. The batches execute in sequence, with independent local
solves running concurrently within each batch. A ratio $\rho>1$ corresponds
to a fractional reduction in online query cost of $1-1/\rho$.

Uncertainty estimates retain the pairing between methods. For fixed-demand
unions we resample workload blocks. For comparisons across independently
drawn graphs we aggregate repetitions within graph before resampling graphs
within family and size strata where applicable. For the transfer study on
one road network, we resample whole regions, keeping their nested subgraphs
and queries together. Those intervals describe variation across regions of
that map. In the schedule and update-order studies, eight prespecified
contrasts use 10,000 paired bootstrap draws with marginal 95\% and
Bonferroni-adjusted 99.375\% intervals for synthetic graphs. The original
union and corridor comparisons report descriptive paired intervals. Their
design supports estimation of effect sizes, while formal multiplicity-adjusted
claims would require a prespecified test.

\section{Experiments}
\label{sec:experiments}

\subsection{Computational experiments on routing and annealing}

\subsubsection{Retaining arcs and selecting regions}

The first experiment tested whether optimizing a set of routes jointly
saves arcs when several demands can share part of a path. We used square
grids of side 4, 6, and 8 with unit or heterogeneous weights. Within each
workload block, all methods received the same demands and candidate paths.
We compared simulated annealing and tabu search with a greedy method that
adds paths according to their incremental arc cost. We measured the number
of retained arcs after verification and repair, first under exact
preservation and then with a permitted stretch of 10\%. A separate
comparison using the Ocean simulated annealer included 30 paired workload
blocks for exact preservation. Repeated solver runs within a block measured
variation in the choice of paths for those same demands.

The corridor experiment used grids of side 32 and 100 and the directed
New York road-distance graph. For a fixed partition and query, we compared
the work-weighted QUBO with corridors initialized by quotient distance or
by the minimum number of regions. Each initializer used the boundary test
in Eq.~\eqref{eq:corridor-certificate}. We counted original-arc
inspections through certification, expansion, and full-graph fallback so
that the comparison charged each method for reaching a verified answer.
In the separate Ocean comparison, six synthetic graphs, three partition
fits, and four common queries per fit gave 72 paired query conditions.
Comparisons between solvers on the same model used identical coefficients
and recovery rules. Comparisons with full-graph and hierarchical searches
included preparation and query time. All annealing measurements were made
on classical processors.

\subsubsection{Constructing reusable routing structures}

Four experiments tested whether QUBO objectives could improve decisions
made before a query. For separator ordering, we compared weights derived
from expected triangle counts with unit separator weights. For landmark
selection, we compared a workload-based objective with farthest-point
selection at the same landmark count. For multiscale covers, we compared
joint selection across scales with independent selection at each scale.
For contraction hierarchies, we compared selecting batches of vertices
with contracting one vertex at a time by a greedy rule. The respective
endpoints were customization triangle relaxations, A* arc inspections,
stored cover entries subject to a local density bound, and query arc
inspections.

An initial comparison used eight grid or sparse-random graphs per task.
The separator procedure was revised after inspection of those cases, which
limits their use to exploration. A subsequent study used 1,650 held-out
directed rings with additional chords, sampled with chord probability
$1/20$ and integer weights from 1 to 9. Table~\ref{tab:experimental-qubos}
gives the allocation by task. Each graph contributed one paired endpoint
after aggregation over repeated runs. A further test of separator weighting
used 200 new ten-vertex graphs satisfying a sparsity criterion identified
in the preceding study. These graphs had at most 23 arcs and admitted a
balanced root separator. The prespecified comparison asked whether separator
weighting reduced customization attempts by at least 5\% on this population.

\subsubsection{Routing through blocks}

The overlay experiment tested how block size and graph topology affect
the cost of replacing internal paths with boundary-to-boundary arcs.
Its 45 graphs comprised bidirectional subgrids, directed rings with
chords, and graphs with densely connected communities. Each family had five
draws at each of 16, 64, and 256 vertices, with directed arc weights from 1
to 9. For each graph we sampled 40 ordered endpoint pairs before inspecting
reachability and used them across all treatments. We formed blocks by
breadth-first traversal with target sizes of 4, 8, and 16 vertices. Boundary
paths were reusable, while endpoint attachments were computed for each query.

The five treatments were full-graph Dijkstra, an exact overlay built with
local Dijkstra searches, an overlay built by simulated annealing, an
overlay built by exhaustive QUBO enumeration where local models had at
most 20 bits, and an annealed overlay followed by a full-graph check.
The annealer used 100, 1,000, or 10,000 sweeps and five seeds. We measured
route availability, stretch of returned routes, final arc scans, and the
work of preparing and querying the overlay. A repeat on the same workload
with the Ocean sampler recorded time spent outside sampling, including
conversion and decoding.

A second comparison extended the exact classical overlay to block targets
of 32 and 64 where the graph size permitted and added the full New York
graph (264,346 vertices and 733,846 directed arcs) with 256 sampled queries.
Each of five measured repetitions rebuilt the overlay after a discarded
warm-up. For each eligible graph and block size, we paired final scans and
complete online time with the direct Dijkstra measurements on the same
queries. Preprocessing time was recorded separately.

\subsubsection{Sampling a local shortest path}

For a fixed local graph and endpoint pair, we varied the encoding and
counted a success when the complete sampled assignment decoded to an
optimal route. An initial comparison used eight encodings and four
analytically motivated variants, with breadth-first search and Dijkstra
as controls. The exploratory and fresh validation sets each contained one
graph per family at each of 4 and 9 vertices, two queries per graph, and
five solver seeds. At a ceiling of 4,096 update proposals, we compared the
sum-of-weights penalty in Eq.~\eqref{eq:local-flow} with the penalty derived
from the breadth-first path witness.

We then tested how the two penalties performed when the budget covered
the entire computation. On 168 fresh graphs at 4, 9, 16, and 25 vertices,
we crossed fixed or coefficient-derived inverse-temperature schedules with
sequential or random variable updates. Each graph supplied two reachable
queries and three repetitions per treatment. The primary endpoint was
strict optimal success within 20\,ms, including construction and decoding.
A transfer experiment used 90 new synthetic graphs at 16, 36, and 64
vertices and three nested subgraphs in each of ten disjoint New York road
regions. At the same deadline, it compared the two flow penalties under
each schedule. At 16,384 proposals, it also compared the original variable
order with three fixed permutations, preserving the initial assignment
under relabeling. The road comparisons used within-subgraph reference
distances and treated each region as a single sampling unit.

\subsection{Theoretical analysis and finite experiments on QUBO matrices}

\subsubsection{Sign structure and exact solution of reduction models}

We examined the integer polynomials from seven illustrative demand-union
models and six corridor models used in the Ocean experiments. For each
nonzero off-diagonal coefficient we recorded the sign and the variables it
coupled. We tested whether all such coefficients were nonpositive, in which
case the QUBO admits an exact minimum-cut solution. For binary variables,
$x_i x_j=(x_i+x_j-|x_i-x_j|)/2$. Multiplication by a negative coefficient
therefore gives a nonnegative cut-edge weight and two unary terms. We
included the constant offset when comparing the minimum energy with that
of the original polynomial.

\subsubsection{Matrices generated during routing optimization}

The routing-structure experiments provided a direct test of what an energy
minimum predicts about an eventual query. We recovered the quadratic models
actually submitted during separator recursion, landmark selection, cover
construction, and contraction, together with the returned assignments.
For the initial eight-graph comparisons, we also examined the spectrum of
the symmetric coefficient matrix obtained by placing half of each pair
coefficient in each off-diagonal position. After scaling by the largest
absolute entry, we used a relative eigenvalue tolerance of $10^{-10}$ to
identify numerical singularity. For the subsequent held-out study we traced
energy-minimizing assignments through the complete routing procedure.
We counted repeated solves separately from distinct numerical
polynomials and identified each polynomial by its ordered
variables, coefficients, and constant term. For each model we measured
coefficient ranges, the number of nonzero interactions, and the degrees
and connected components of the interaction graph. For models with at most
16 bits, exhaustive enumeration gave a reference minimum against which to
score the raw assignment before repair or routing.

For a fixed matrix, we varied a feasible decision, completed the remaining
routing procedure, and compared energy with the resulting work. On each of
the 50 landmark graphs we evaluated all 45 candidate pairs on both training
and held-out queries, resolving energy ties lexicographically. For recursive
decisions each candidate required completing the hierarchy before evaluation.
We evaluated all 68 feasible root assignments per separator objective on
one graph and all 11 feasible first contraction batches on another, keeping
the continuation rule fixed within each comparison.

For covers, we tested whether the selected entries intersected every required
shortest path, including tied paths. Repeating this check before and after
pruning allowed us to locate any loss of coverage. We checked the answers
returned by the routing algorithm separately.

\subsubsection{Interaction structure of local path models}

In the synthetic overlay experiment we collected the matrices submitted
for local path searches. For each distinct matrix we measured the number of
variables, nonzero couplings, coefficient ranges, and bounds on the treewidth
of the interaction graph. We used the incidence matrix in
Eq.~\eqref{eq:local-flow} to determine which pairs of arc variables interact.

To isolate the effect of representation, we selected nine local endpoint
pairs from 256-vertex parent graphs, one per graph family and block target.
We constructed eight encodings for each pair, holding the local arcs,
weights, and endpoints fixed. We compared matrix dimensions, interaction
patterns, and the largest absolute coefficient within each set of eight.
A separate extraction
from the exact-overlay comparison preserved 671 induced blocks and 5,804
endpoint-pair records for subsequent study. The fresh local-solver
validation used independently drawn inputs. A later comparison on 96 paired
queries examined the cost of tightening the validity penalty in the
domain-wall encoding. We measured the reduction in coefficient magnitude
and the time needed to construct the tighter bound for each query.

\subsubsection{Ground states and the energy landscape}

For the eight local encodings we asked whether every ground state decodes
to a shortest route and whether each shortest route has a representative.
The proofs use positive integer weights, distinct reachable endpoints,
and loop-free directed multigraphs. Arc encodings distinguish parallel
arcs, whereas position encodings distinguish vertex sequences. We used
assignments attaining the flow-penalty threshold, minimized over auxiliary
bits to compare degree penalties, and bounded the energy of invalid
domain-wall configurations. Sixteen explicit examples tested the limits
of these arguments, including how the penalty threshold changes when an
endpoint conservation equation is omitted.

We checked the implemented polynomials separately on all 738 labeled
two- and three-vertex graph descriptions with arc weights in $\{1,2\}$ or
absent arcs. There were 3,576 reachable ordered queries and 816 unreachable
pairs. For eligible models of at most 20 bits, we enumerated the binary
assignments, compared matrix energies including offsets with independently
evaluated polynomials, and decoded every ground state. On a systematic
subsample augmented with cases involving ties, parallel arcs, cycles, and
branching, we also examined single-bit moves. For models of at most ten
bits we computed the smallest energy barrier along a sequence of flips to
a ground state, allowing moves along plateaus. These finite checks probe
features of the landscape that remain free to vary even when two encodings
represent exactly the same optimal routes.

\section{Results}
\label{sec:results}

\subsection{Verified reduction and complete query cost}

The demand unions preserved the full-graph distance for every tested demand
at zero tolerance, and the certified corridor searches returned exact final
answers. Some queries required repair of the proposed union, corridor
expansion, or a full-graph fallback to reach that outcome. For fixed demands, simulated
annealing retained 5.6\% more arcs than greedy selection after verification
(paired ratio 1.056, 95\% interval 1.049--1.065). Allowing paths up to
10\% longer reduced the annealed union by only 1.3\% relative to exact
preservation. For new queries, the work-weighted corridor QUBO required
31.9\% more cumulative arc inspections than initialization by quotient
distance (ratio 1.319, 95\% interval 1.289--1.350).

\begin{figure}[htbp]
\centering
\includegraphics[width=\linewidth]{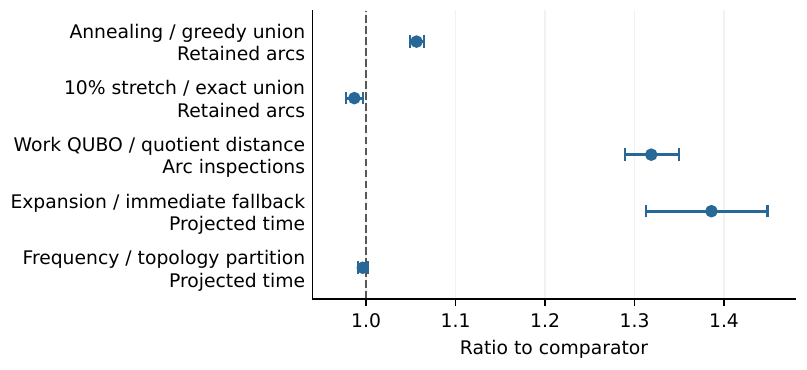}
\caption{Paired effects for verified demand unions and corridors. Points
give ratios of the first method to the second for the outcome named in
each row. Bars give descriptive 95\% bootstrap intervals. Values below
one favor the first method. Time includes setup and the projected cost of
1,000 queries. The expansion comparison covers the New York and unit-grid
conditions specified for that comparison.}
\label{fig:verified-reduction}
\end{figure}

Recovery policy accounted for part of the corridor cost. Expanding an
uncertified corridor before fallback increased projected time at 1,000
queries by 38.6\% relative to immediate fallback
(Fig.~\ref{fig:verified-reduction}). On the six synthetic graphs in the
separate Ocean comparison, changing the solver on the same corridor QUBO
changed the amount of subsequent search work. Simulated annealing required
23.8\% fewer
cumulative inspections than tabu search (paired ratio 0.762, exploratory
95\% interval 0.680--0.862). Full-graph fallback occurred in 36 of 360
annealing invocations and 105 of 360 tabu invocations. Quotient-distance
initialization still had the lowest median inspection count of these three
methods.

\subsection{Final-search savings and the cost of using an overlay}

\begin{figure}[htbp]
\centering
\includegraphics[width=\linewidth]{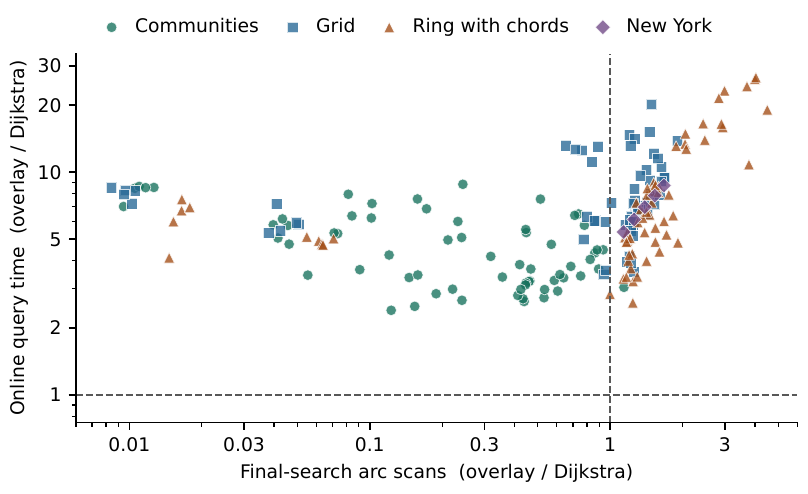}
\caption{Exact block overlay versus full-graph Dijkstra. Each point compares
one graph at one block size on identical queries. A point left of the
vertical line denotes fewer final arc scans. A point above the horizontal
line denotes a slower online query, including endpoint attachment, assembly,
search, and reconstruction. Both axes are logarithmic. Several block sizes
were evaluated on each graph.}
\label{fig:overlay-cost}
\end{figure}

Exact local paths reduced the number of arcs scanned by the final Dijkstra
search in 97 of 200 paired graph and block-size comparisons. In all 200,
the complete online query took longer than direct Dijkstra
(Fig.~\ref{fig:overlay-cost}). The largest contrast at 256 vertices
occurred on graphs with community structure. At a target block size of 64,
the exact overlay used 8.6\% of the baseline final scans but took 6.04
times as long per query. On the full New York graph at the same block
size, final scans increased to 1.67 times the baseline and query time to
8.70 times the baseline. Assembling the query overlay there took 2.13\,s,
or 80.7\% of its online time. Direct Dijkstra answered the complete query
in 0.304\,s on average.

Replacing exact local transitions with annealed paths could leave the final
scan count almost unchanged while substantially degrading route quality.
On 256-vertex community graphs with blocks of 16, the exact and annealed
overlays used 18.56\% and 18.59\% of the baseline scans, respectively.
After 10,000 sweeps the annealed overlay found routes for all reachable
queries in this condition, but only 12.1\% had minimum cost. At block
size four, the same sweep budget gave exact answers for every tested
reachable query across the three families and sizes. Checking each
annealed route with full-graph Dijkstra restored exact answers at larger
block sizes and incurred the full baseline search cost on every query.

\subsubsection{Estimated query speedup with a prepared overlay}

We evaluate Eq.~\eqref{eq:parallel-speedup} on a modeled graph of 16,384 vertices
with $k$ vertices and two boundary vertices per block, $2k$ internal arcs
per block, and one outgoing interblock arc per boundary vertex. Use the
search cost proxy $d(v,e)=(v+e)\log_2 v$ and charge one unit per arc read
or write outside search. Let $K=16{,}384/k$. With five endpoint connections
and a stored static overlay, the common query cost in these units is
\[
 C=d(2K+2,4K+5)+2\lceil k\log_2(K)/2\rceil+5(1+k/2).
\]
The terms count overlay search, route reconstruction, and attachment
materialization, respectively. As a classical comparator, we allow the
forward and reverse local Dijkstra searches to run concurrently on
prebuilt adjacency lists, giving $D=C+d(k,2k)$. This comparison credits
both methods with the same partition, precomputation, and opportunity to
parallelize independent searches.

At $k=64$, the estimated query speedup is $1.072\times$ relative to the
parallel classical overlay, corresponding to a 6.7\% reduction in online
cost. The shared work costs 14,573 proxy units per query. The classical
local searches add 1,152 units on the critical path. If all five local
problems fit in one quantum batch and its complete service costs 100 units,
the respective query totals are 15,725 and 14,673 units (rounded).
Serving the five problems sequentially at 100 units each gives a speedup of
$1.043\times$ (4.1\% lower cost). At $k=32$, even with all five problems in
one batch, the estimated speedup is $1.012\times$ (1.2\% lower cost),
because the global overlay search accounts for a larger fraction of the
query. These estimates assume that the batch returns local paths of the
required quality. Batch capacity and service cost at that quality remain
unmeasured on quantum hardware.

The measured New York phases give a separate comparison with direct
Dijkstra. At block size 64, direct Dijkstra took 303.7\,ms per query and
the final overlay search alone took 298.2\,ms. Holding the latter time
fixed and assigning zero cost to every other online phase gives an
optimistic upper bound of $1.018\times$ on the speedup for that measured
breadth-first partition. A fixed grid could produce a different overlay
and hence a different final search cost.

\subsection{QUBO energy and the quality of routing structures}

\subsubsection{Sign structure and certified energy minima}

\begin{figure}[htbp]
\centering
\includegraphics[width=\linewidth]{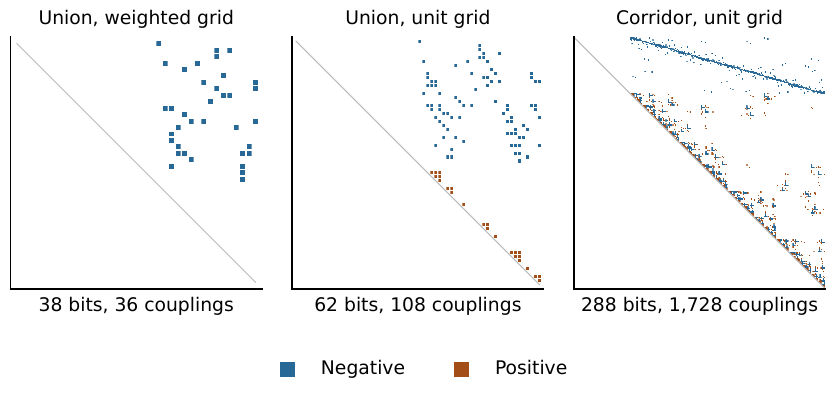}
\caption{Signs of the off-diagonal coefficients in three reduction QUBOs.
Each mark represents one interacting variable pair in the stored upper
triangle. The union examples use $4\times4$ grids with exact preservation.
The corridor example uses a $16\times16$ grid. Interaction densities are
5.1\%, 5.7\%, and 4.2\% from left to right. Only the first example has
exclusively negative couplings.}
\label{fig:coupling-signs}
\end{figure}

Three of the seven demand-union models admitted exact solution by a
minimum cut. They came from the weighted grids of side 4, 6, and 8 at
zero stretch tolerance. The certified minimum energies, including the
offset, were 22, 37, and 54, respectively. In these examples the only
couplings linked a candidate path to the arcs it required. The unit-grid
examples also coupled competing paths for the same demand, introducing
positive coefficients through the one-path constraint in
Eq.~\eqref{eq:demand-union}. All six corridor models had mixed signs.
Figure~\ref{fig:coupling-signs} illustrates why sparsity alone does
not identify the models amenable to this direct cut construction.

In the initial routing-structure study, 690 solver calls used 240
distinct matrices. The eight landmark matrices each had ten variables
with positive couplings between every pair. The cover matrices had a
median interaction density of only 12.9\%, but reached 95 variables.
The symmetric coefficient matrix was numerically singular in 12 of 24
cover models and 3 of 42 separator models. This spectral property
describes the continuous quadratic form. To assess the discrete solver
directly, exhaustive enumeration certified the minima of 201 matrices
used in 495 calls. Eight returned assignments missed the certified
minimum, all in the 80 landmark calls. The remaining 195 calls involved
larger models whose minima were uncertified.

\subsubsection{Energy minima and subsequent routing work}

Across the 1,650 held-out graphs, none of the four routing structures met
the prespecified criterion for a 5\% improvement. The ratio of customization
attempts with triangle-weighted separators to attempts with size-only
separators was 0.930 (simultaneous interval 0.892--0.969). On the
independent sample of 200 graphs from the restricted sparse class, the
ratio was 0.934 (95\% interval 0.911--0.958). Both intervals span the
practical threshold of 0.95. In the independent sample, 89 graphs produced
identical counts under both objectives. For landmarks, the held-out
inspection ratio was 0.952 (simultaneous interval 0.891--1.017). For
contraction batches it was 1.002 (0.957--1.050).

\begin{table}[htbp]
\centering
\caption{QUBOs generated in the held-out routing-structure experiment.
Repeated solves on a graph could submit the same polynomial. The last
column gives the observed number of binary variables per model.}
\label{tab:experimental-qubos}
\small
\begin{tabular}{@{}lrrrr@{}}
\hline
Routing decision & Graphs & Solves & Distinct polynomials & Variables\\
\hline
Separator ordering  & 200   & 6,475  & 628   & 13--36\\
Landmark selection  & 50    & 500    & 50    & 10\\
Multiscale covers   & 1,200 & 18,000 & 3,600 & 14--176\\
Contraction batches & 200   & 7,552  & 2,232 & 1--5\\
\hline
\end{tabular}
\end{table}

The experiments generated 32,527 solve observations across the four tasks
(Table~\ref{tab:experimental-qubos}). Exact energy checks on eligible
models identified a limitation of the contraction objective. All 7,552 local contraction solves attained their global
QUBO minimum. Yet on the graph for which every feasible first batch was
evaluated, the unique minimum-energy batch led to 91 query arc inspections.
Another feasible first batch led to 65 under the same continuation rule.
Among landmark choices, the energy-minimizing pair with the specified
tie rule failed to minimize held-out inspections on 46 of 50 graphs. The
median within-graph rank correlation between energy and held-out inspections
was 0.437. Thus reducing solver error alone would still leave a discrepancy
between the optimized objective and subsequent search work.

For multiscale covers, joint selection produced 0.991 times as many stored
entries as independent selection. Local density exceeded that of the control
on 370 of 1,200 graphs. More fundamentally, all 12,000 final covers omitted
at least one required shortest path. In 10,693 cases the decoded selection
covered every required path before pruning. The loss of coverage in those
cases arose during pruning. Fallback search returned correct routes, while
the entry counts measured structures with incomplete coverage.

\subsection{Matrix structure and recovery of optimal local paths}

The 35,165 distinct local flow matrices from the synthetic overlays ranged
from 1 to 47 variables and from 0 to 227 nonzero quadratic interactions.
The incidence formulation explains how interactions accumulate at high-degree
vertices. In Eq.~\eqref{eq:local-flow} the coefficient for two distinct arcs
is $Q_{ef}=2A\sum_v B_{ve}B_{vf}$. Two arcs entering the same vertex
contribute $+2A$ there, while an entering and a leaving arc contribute
$-2A$. On the nine local problems with eight paired encodings, tightening
the flow penalty preserved the interaction pattern and reduced the largest
absolute coefficient to a median of 8.64\% of its original value.
Changing to a one-hot position encoding increased the observed range from
3--42 to 8--224 variables and from 2--177 to 27--6,165 interactions.

All eight formulations represented the optimal route cost under their
stated penalty bounds. For full flow balance the sharp threshold is
$A>C^*/2$, where $C^*$ is the local shortest-path cost. Omitting the
source or target balance equation preserves feasible flows but reduces
the squared residual of the empty flow from two to one, raising the sharp
threshold to $A>C^*$ for the unmodified traversal costs. The finite checks
and candidate screen evaluated 11,974,374 assignments. Every enumerated
baseline ground state met its decoding criterion.

Two encodings can represent the same shortest routes and assign different
energies to a branching path candidate. If $d$ outgoing arcs are selected
at one vertex, the pairwise penalty is $\binom{d}{2}$ and the slack
penalty is $\min_{z\in\{0,1\}}(d+z-1)^2$. Both vanish at $d=0,1$ and
equal one at $d=2$. At $d=3$ their contributions are three and four,
respectively. On the six fresh
validation graphs, the pairwise encoding returned a strictly valid optimum
in 53 of 60 trials at 4,096 proposals, compared with 42 of 60 for the
slack encoding. The paired difference was 18.3 percentage points
(descriptive 95\% graph-bootstrap interval 5.0--33.3). The comparison
captures the combined change in energy penalties and number of variables.

At 4,096 update proposals, replacing the sum-of-weights penalty with the
breadth-first witness penalty increased strictly optimal returns from
17 of 60 to 51 of 60 on six fresh graphs. On the subsequent 168-graph
sample, the tighter penalty yielded 848 optimal returns in 1,008 trials
within 20\,ms, compared with 361 for the original penalty under the same
fixed schedule and sequential updates. The paired gain was 48.31 percentage
points (multiplicity-adjusted 99.375\% interval 43.85--52.65). Random
variable selection reduced success for the tighter penalty by 3.82 points
relative to sequential updates, averaged over schedules. Dijkstra returned
an optimum within the deadline in every primary control trial, with a mean
service time of 0.0179\,ms.

\begin{figure}[htbp]
\centering
\includegraphics[width=\linewidth]{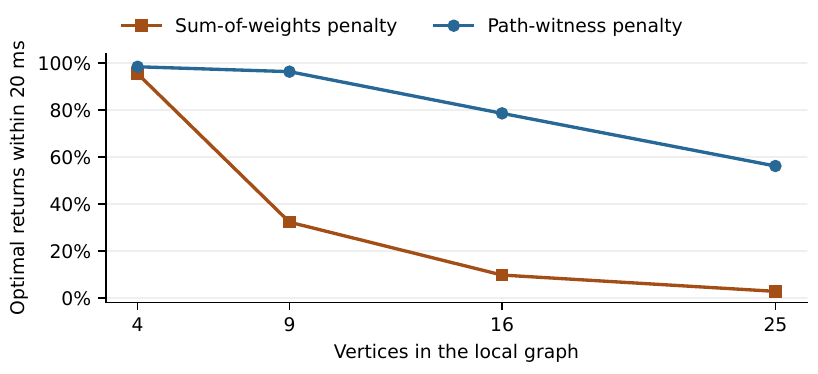}
\caption{Strictly optimal returns within the complete 20\,ms deadline
for two penalties applied to the same arc-flow encoding. Each point
summarizes 42 graphs at one size, weighting three graph families and four
schedule and update combinations equally. These descriptive rates average
over policies. The paired 48.31-point contrast in the text uses the fixed
schedule and sequential updates.}
\label{fig:flow-penalty-success}
\end{figure}

The gain varied with graph size and population. At four vertices both
penalties returned optima in over 95\% of trials when averaged over
policies. At 25 vertices the corresponding rates were 2.8\% for the
original penalty and 56.2\% for the witness penalty
(Fig.~\ref{fig:flow-penalty-success}). On the 90 larger synthetic graphs
the gain under sequential updates persisted under both the fixed and
coefficient-derived schedules (45.56 and 44.81 percentage points).
In the ten New York regions the gains were 0.56 points under the fixed
schedule (descriptive 95\% interval $-4.44$ to 5.56) and 28.33 points
under the coefficient-derived schedule (22.22 to 35.00). Even in the
latter condition the tighter penalty yielded only 52 on-time optima in
180 trials.

The cost of constructing a tighter bound also depended on the encoding.
Grouping terms in the domain-wall penalty reduced its coefficient on all
96 paired diagnostic queries. On 82 of them, constructing that bound alone
took longer than the 20\,ms deadline on average. The observed gains for
the arc-flow penalty therefore concern recovery of a local optimum by CPU
annealing. Whether they translate into faster verified overlay queries
remains open.

\section{Discussion and Conclusion}
\label{sec:discussion-conclusion}

The value of a reduction depends on how it will be used. When demands are
known in advance, the union of validated paths gives a direct preservation
guarantee, and a greedy choice of paths can already exploit shared arcs.
For an incoming query, the reducer must also establish that the omitted
part of the graph contains no shorter route. In our experiments, a
corridor initialized by quotient distance required fewer arc inspections
through certification than the work-weighted QUBO. The matrix analysis
helps explain why improving a solver can have limited effect. In the
contraction experiment, even an exactly minimized local objective could
select a first batch that increased the work of subsequent queries. This
suggests evaluating a proposed objective against the cost of the routing
decisions it induces before investing in a faster optimizer.

For block overlays, the arrangement of boundary vertices helps explain
the differences between graph families. A block with few connections to its
neighbors can hide many internal arcs behind a few exact transitions.
Other partitions require transitions between many ordered pairs of
boundary vertices. On the New York graph, increasing the target block
size reduced the fraction of vertices on a boundary, yet increased the
number of overlay arcs. Rebuilding those arcs for each query consumed much
of the measured online time. Reusing a static overlay addresses that cost
when the graph and weights remain fixed. The cost of attaching new endpoints
and the accuracy of the local paths still determine how much of the
final-search saving reaches the complete query.

We used local CPU implementations of classical simulated annealing,
including the Ocean sampler, to evaluate QUBO formulations intended for
quantum annealing. The local path experiments show what can be gained by
changing the formulation alone. Tightening the flow penalty preserved the
optimal paths and the interaction graph while increasing the frequency of
optimal returns within a fixed deadline. Transfer depended on the workload
and the temperature schedule. Even under the more favorable schedule on
the New York subgraphs, the tighter formulation returned a local optimum
on time in only 52 of 180 trials. A route may depend on several such local solutions,
and the time spent obtaining and verifying them belongs in any estimate
of the cost of an exact query. The road results concern regions of one
map, and the CPU measurements leave the cost and quality of a physical
annealing batch open.

\paragraph{Conclusion.}
We studied how the subgraph retained for a demand, the boundary paths
reused across queries, and the encoding of a local path each affect the
cost of returning a verified route. The clearest opportunity for
parallel execution arises when a prepared overlay leaves several independent
endpoint problems on the query's critical path. For the modeled graph of
16,384 vertices with 64 vertices and two boundary vertices per block,
solving five endpoint problems in one batch at an assumed cost of 100
proxy units gives an estimated query speedup of $1.072\times$ relative
to the parallel classical overlay. This corresponds to a 6.7\% reduction
in modeled online cost at matched route quality. Testing that estimate on
a quantum annealer requires measuring the complete batch latency at the
accuracy needed to return a verified route.

Our next application is the FLS multigraph of an information architecture.
Its vertices represent modules, and a directed interaction requires
compatible physical, syntactic, and semantic links enabled by the current
access controls. We plan to apply the reducer to queries about admissible
information flow between specified modules. Each retained transition must
preserve the identities and direction of its constituent links so that a
returned route can be checked against the original architecture. We will
compare reachability, minimum admissible path cost, and verified query
latency with an exact search under the same access rules. This will test
whether the structure of FLS workloads permits enough reuse to realize
the modeled query savings.

\bibliographystyle{plain}
\bibliography{references}

\end{document}